\documentclass[PAPER, cernpreprint, atlasdraft=false, texlive=2023, USenglish, texmf, orcidlogo]{atlasdoc}
\usepackage[subfigure=true]{atlaspackage}
\usepackage{atlasbiblatex}

\usepackage{atlasphysics}
\usepackage{atlasmisc}
\graphicspath{{logos/}{figures/}}

\usepackage{ANA-HION-2022-10-PAPER-defs}

\AtlasTitle{Observation of $t\bar{t}$ production in Pb+Pb collisions at $\sqrt{s_\mathrm{NN}}=5.02$~\tev\ with the ATLAS detector}

\AtlasVersion{2.6}

\AtlasAbstract{%
Top-quark pair production is observed in lead-lead~(Pb+Pb) collisions at $\sqrt{s_\mathrm{NN}}=5.02 $~\ensuremath{\text{TeV}}\xspace
at the Large Hadron Collider with the ATLAS detector. The data sample was recorded in 2015 and 2018, amounting to an integrated luminosity of 1.9~nb$^{-1}$. Events with exactly one electron and one muon and at least two jets are selected.
Top-quark pair production is measured with an observed~(expected) significance of 5.0~(4.1) standard deviations. The measured top-quark pair production cross-section is
$\sigma_{t\bar{t}} = 3.6\;^{+1.0}_{-0.9}\;\mathrm{(stat.)}\;^{+0.8}_{-0.5}\;\mathrm{(syst.)}
~\mathrm{\mu b}$, with a total relative uncertainty of 31\%, and is consistent with theoretical predictions using a range of different nuclear parton distribution functions.
The observation of this process consolidates the evidence of the existence of all quark flavors in the pre-equilibrium stage of the quark-gluon plasma at very high energy densities, similar to the conditions present in the early universe.
}

\AtlasRefCode{HION-2022-10}

\PreprintIdNumber{CERN-EP-2024-282}

\AtlasJournal{Phys. Rev. Lett.}
\AtlasJournalRef{\PRL 134 (2025) 142301}
\AtlasDOI{10.1103/PhysRevLett.134.142301}

\hypersetup{pdftitle={ATLAS document},pdfauthor={The ATLAS Collaboration}}

\begin{document}

\maketitle

Heavy-ion~(HI) collisions at the Large Hadron Collider~(LHC) recreate the quark-gluon plasma~(QGP) in a laboratory setting.
This exotic state of matter, consisting of deconfined quarks and gluons, is believed to have existed in the early universe.
However, the QGP is short-lived, with a lifetime of the order of 10 fm/$c$ ($\approx 10^{-23}$ s), making direct measurement impossible~\cite{Busza:2018rrf}.
Instead, particles produced in the collisions and passing through the QGP are used as probes, providing essential insights
into QGP properties, and therefore improving our understanding of the development of the early universe and Quantum Chromodynamics~(QCD)~\cite{Busza:2018rrf, iancu2012qcd, Ratti:2018ksb} under extreme conditions.

Top quarks produced in HI collisions are expected to offer valuable new experimental insights~\cite{dEnterria:2015mgr}.
Of particular interest, is the study of hadronically decaying $W$ bosons from top-quark decays, which can yield unique information about the time structure of the QGP~\cite{Apolinario:2017sob}.
This is made possible by the time delay between the decay of the top quark and the subsequent decay of the $W$ boson, offering insights unattainable by any other current experimental method.
Additionally, top-quark measurements in HI collisions may provide constraints on nuclear modifications to the parton distribution functions (nPDFs)~\cite{Khanpour:2016pph, Ethier:2020way}, especially in kinematic regions that are poorly constrained by other measurements.

At the LHC, top quarks are predominantly produced in top quark--antiquark ($t\bar{t}$) pairs, with pair production being more common than single top-quark production~\cite{Giammanco:2017xyn}.
The top quark is highly unstable, almost always decaying into a $W$ boson and a $b$-quark. Its lifetime is shorter than the timescale to form hadrons, meaning it decays before hadrons can form.
The $W$ boson then decays  either leptonically ($W \rightarrow \ell\nu$) or hadronically ($W \rightarrow q\bar{q}^\prime$). In LHC collisions, top-quark production is primarily the result of gluon--gluon fusion.
The $t\bar{t}$ production cross section is calculated at next-to-next-to-leading-order~(NNLO) and next-to-next-to-leading-logarithmic~(NNLL) precision in QCD~\cite{Czakon:2013goa}.
With large integrated luminosity data samples from proton--lead (\pPb) and lead--lead (\PbPb) collisions, it becomes experimentally feasible to reconstruct $t\bar{t}$ events in the \ljets{} ($t\bar{t} \rightarrow WbW\bar{b} \rightarrow \ell\nu_\ell bq\bar{q}^\prime\bar{b}$) and dilepton ($t\bar{t} \rightarrow WbW\bar{b} \rightarrow \ell\nu_\ell b\ell\bar{\nu}\ell\bar{b}$) channels~\cite{dEnterria:2017jyt}.
The ATLAS and CMS collaborations have observed $t\bar{t}$ production in the \ljets{} and dilepton channels in \pPb{} collisions at $\sqrt{s_\mathrm{NN}}=8.16 $~\TeV~\cite{CMS-HIN-17-002, TOPQ-2023-32}, and CMS has provided evidence of $t\bar{t}$ production in the dilepton channel with a significance of $4.0\;\sigma$ using \PbPb{} collisions at $\sqrt{s_\mathrm{NN}}=5.02 $~\TeV\cite{CMS-HIN-19-001}.

The first observation is presented of $t\bar{t}$ production in \PbPb{} collisions at $\sqrt{s_\mathrm{NN}}=5.02$~\TeV using data from the ATLAS experiment.
Top-quark pair events are selected in the electron-muon ($e\mu$) decay channel, $t\bar{t} \rightarrow e\nu_e \mu\nu_\mu b \bar{b}$, as this channel has the least background contamination.
The electron-electron ($ee$) and muon-muon ($\mu\mu$) channels, which are dominated by $Z \rightarrow \ell\ell$ production, are used to derive and validate detector response corrections.
The data sample was collected in 2015 and 2018, corresponding to a total integrated luminosity of 1.9~nb$^{-1}$.
No requirements are imposed on $b$-quark-initiated jets, allowing it to serve as a basis for calibrating heavy-flavor jets in HI collisions. As such, this work represents a significant milestone for future studies of hadronically decaying top quarks in a QGP.

The ATLAS experiment~\cite{PERF-2007-01} at the LHC is a multipurpose particle detector
with a forward--backward symmetric cylindrical geometry and a near \(4\pi\) coverage in
solid angle.\footnote{ATLAS uses a right-handed coordinate system with its origin at the nominal interaction point (IP)
in the center of the detector and the \(z\)-axis along the beam pipe.
The \(x\)-axis points from the IP to the center of the LHC ring,
and the \(y\)-axis points upwards.
Polar coordinates \((r,\phi)\) are used in the transverse plane,
\(\phi\) being the azimuthal angle around the \(z\)-axis.
The pseudorapidity is defined in terms of the polar angle \(\theta\) as \(\eta = -\ln \tan(\theta/2)\) and is equal to the rapidity
$ y = \frac{1}{2} \ln \left( \frac{E + p_z}{E - p_z} \right) $ in the relativistic limit.
Angular distance is measured in units of \(\Delta R \equiv \sqrt{(\Delta y)^{2} + (\Delta\phi)^{2}}\).}
It consists of an inner tracking detector~(ID) surrounded by a thin superconducting solenoid
providing a \qty{2}{\tesla} axial magnetic field, electromagnetic and hadronic calorimeters, and a muon spectrometer~(MS).
The ID covers the pseudorapidity range \(|\eta| < 2.5\).
It consists of silicon pixel, silicon microstrip, and transition radiation tracking detectors.
Lead/liquid-argon (LAr) sampling calorimeters provide electromagnetic (EM) energy measurements
with high granularity within the region \(|\eta|< 3.2\).
A steel/scintillator-tile hadronic calorimeter covers the central pseudorapidity range (\(|\eta| < 1.7\)).
The endcap and forward regions are instrumented with LAr calorimeters
for EM and hadronic energy measurements up to \(|\eta| = 4.9\).
The MS surrounds the calorimeters and is based on
three large superconducting air-core toroidal magnets with eight coils each.
The field integral of the toroids ranges between \num{2.0} and \qty{6.0}{\tesla\metre}
across most of the detector.
The MS includes a system of precision tracking chambers up to \(|\eta| = 2.7\) and fast detectors for triggering up to \(|\eta| = 2.4\).
The luminosity is measured mainly by the LUCID--2~\cite{LUCID2} detector, which is located close to the beampipe.
A two-level trigger system is used to select events~\cite{TRIG-2016-01}.
The first-level trigger is implemented in hardware and uses a subset of the detector information
to accept events at a rate below \qty{100}{\kHz}.
This is followed by a software-based trigger that
reduces the accepted event rate to \qty{1}{\kHz} on average
depending on the data-taking conditions.
A software suite~\cite{SOFT-2022-02} is used in data simulation, in the reconstruction
and analysis of real and simulated data, in detector operations, and in the trigger and data acquisition
systems of the experiment.

The centrality of Pb+Pb events characterizes the degree of geometric overlap of the colliding nuclei. It is defined using $\sum E_{T}$, the total transverse energy measured
in the Forward Calorimeter (FCal)~\cite{HION-2018-06, HION-2016-07}, which covers the $3.1<|\eta|<4.9$ range. Events are categorized into centrality intervals using a Glauber model~\cite{Miller:2007ri} parameterization of the $\sum E_{T}$ distribution. Each interval represents a range in $\sum E_{T}$, starting at 0\% for the most central collisions and ending at 100\% for the most peripheral collisions.

Events are selected using single-electron or single-muon triggers with a minimum transverse momentum~($\pT$) threshold of 15 and 8~\GeV, respectively~\cite{TRIG-2018-05,TRIG-2018-01, TRIG-2019-03}.
The events are required to have exactly one reconstructed vertex built from at least two good-quality charged-particle tracks with $\pT \geq 0.5$~\GeV~\cite{ATL-PHYS-PUB-2015-026}.
Events are selected in the 0--80\% centrality interval to avoid contributions from photon-induced processes. This requirement has a minimal effect on the signal yield as the contribution to the \ttbar\ cross section from the 80--100\% centrality range is estimated to be below 1\%.
The average number of hadronic interactions per bunch crossing was $1.8 \times 10^{-3}$ and $2.6 \times 10^{-3}$
for 2015 and 2018, respectively.

Electron candidates are reconstructed from a localized cluster of energy deposits in the EM calorimeter, matched to a track in the ID\@. Candidates must originate from the primary vertex with the transverse impact parameter significance of $|d_{0}|/\sigma_{d_{0}} < 5$ and the longitudinal impact parameter of $ |\Delta z_{0} \sin\theta| < 0.5$~mm.
They are further required to satisfy the ``Loose'' likelihood-based identification criteria~\cite{EGAM-2018-01} and must have $\pt > 18$~\GeV and $|\eta| < 2.47$. The electron candidates are required to be isolated using track- and calorimeter-based isolation requirements~\cite{HION-2018-25}.

Muon candidates are reconstructed by matching tracks reconstructed in the ID and the MS\@. Candidates are required to originate from the primary vertex with $|d_{0}|/\sigma_{d_{0}} < 3$ and $ |\Delta z_{0} \sin\theta| < 0.5$~mm. They must satisfy ``Loose'' identification criteria~\cite{MUON-2018-03} and are further required to have $\pt > 15$~\GeV and $|\eta| < 2.5$.
The muon candidates must also satisfy track- and calorimeter-based isolation requirements~\cite{HION-2018-25, MUON-2018-03} %

Jets are reconstructed from calorimeter energy deposits~\cite{HION-2017-06} using the anti-$k_t$ algorithm~\cite{Cacciari:2008gp,Fastjet} with a radius parameter $R = 0.4$, and are required to have $\pt > 35$~\GeV and $|\eta| < 2.5$.
Jet kinematics are corrected event-by-event for the contribution from underlying event~(UE) particles, and are calibrated using simulations of the calorimeter response~\cite{JETM-2018-05} and in situ measurements of the absolute energy scale.
In situ measurements are carried out in proton--proton~($pp$) collisions and cross-calibrated to the \PbPb\ data using the method described in Ref.~\cite{ATLAS-CONF-2015-016}, providing an initial jet calibration. Since jet modeling influences calculations of the missing transverse momentum~($\vec{p}_{\mathrm{T}}^{\mathrm{miss}}$, with magnitude $E_\mathrm{T}^\mathrm{miss}$), no requirement on $E_\mathrm{T}^\mathrm{miss}$ is imposed in the analysis.
No $b$-tagging requirements are imposed on jets.

Samples of Monte Carlo~(MC) simulated events are used to develop the analysis procedures, evaluate signal and background contributions, estimate signal efficiencies, and compare the predicted distributions with data. All samples are processed using the full ATLAS detector simulation~\cite{SOFT-2010-01} based on the \geant4 framework~\cite{GEANT4:2002zbu}.
The simulation samples are produced using event-generator configurations similar to those developed for $pp$ analyses~\cite{TOPQ-2020-02,TOPQ-2018-40}.
The nominal simulated $t\bar{t}$ sample is produced using the next-to-leading-order~(NLO) event generator \POWHEGBOX{ v2}~\cite{Alioli:2010xd} with the NNPDF3.0NLO parton distribution function~(PDF) set~\cite{Ball:2014uwa}, where the top-quark mass~($m_t$) is set to 172.5~\GeV.
It is interfaced with \pythia8~\cite{Sjostrand:2014zea} with the NNPDF2.3LO PDF set~\cite{Ball:2012cx} and the A14 set of tuned parameters~\cite{ATL-PHYS-PUB-2014-021} for the parton-shower and hadronization modeling.

Alternative $t\bar{t}$ simulation samples are generated to assess systematic uncertainties related to the signal modeling. In all the top-quark samples, $m_t$ is set to 172.5~\GeV, and the \evtgen{} program~\cite{Lange:2001uf} is used for the decays of $b$- and $c$-flavored hadrons.
One alternative sample uses the \POWHEGBOX{ v2} MC generator with the \herwig~v7.2 parton-shower and hadronization model~\cite{Bahr:2008pv,Bellm:2015jjp}, employing the H7.2-Default set of tuned parameters~\cite{Bellm:2015jjp,Bellm:2019zci}.
Another \POWHEGBOX{ v2}+\pythia8 sample is used to assess the uncertainty in the matrix element and parton
shower matching. Uncertainties in the amount of parton-shower radiation are evaluated by setting the $h_\mathrm{damp}$ parameter, which controls the \pt\ of the first additional gluon emission beyond the leading-order Feynman diagram in the parton shower, to $3 m_t$ instead of the nominal value of $1.5 m_t$.
All signal samples are normalized using a $K$-factor of $K=1.149$ to the cross section calculated at NNLO+NNLL precision in QCD using the \toppp~v2 program~\cite{Czakon:2011xx} with the NNPDF3.0NLO PDF set.

The background processes evaluated using MC simulation are from $Z$ bosons produced  with jets~($Z$+jets), single-top quark~($tW$) and diboson~($VV$) production.
The $Z$+jets events are simulated at NLO precision in QCD, using the \POWHEGBOX{ v1}~\cite{Oleari:2010} generator interfaced with \pythia8, using the AZNLO CTEQ6L1 set of tuned parameters.
The $tW$ associated production processes for single-top quarks are simulated at NLO precision in QCD, using the \POWHEGBOX{ v2}~\cite{Frederix:2012dh,Re:2010bp} generator with the NNPDF3.0NLO PDF set, and using \pythia8 with the A14 NNPDF2.3LO set of tuned parameters.
The diagram-removal scheme is used to treat the interference with $t\bar{t}$ production~\cite{Frixione:2008yi}. The $VV$ backgrounds are simulated at NLO precision in QCD, using the \sherpa~v2.2.14 generator~\cite{Bothmann:2019yzt} with the NNPDF3.0NNLO PDF set~\cite{Ball:2014uwa}.

Simulated signal and background samples are produced separately for four isospin configurations (proton--proton, proton--neutron, neutron--proton and neutron--neutron) and embedded into \hijing~\cite{hijing} minimum-bias \PbPb{} events for UE modeling.
The $Z$+jets samples are produced only for the proton--proton configuration, as no significant differences between isospin combinations  are expected in the cross section. The events are processed using the same reconstruction and analysis chain as the data.
The samples are reweighted on an event-by-event basis to match the $\sum E_{T}$ distribution in Pb+Pb data with a looser selection of at least one lepton~(electron or muon) with a \pT\ above 15~\GeV~(10~\GeV) for electrons~(muons).

To correct for residual differences between data and simulation in the detector response, dedicated corrections are applied accounting for energy and momentum scale and resolution, and lepton efficiencies.
Corrections for electrons are based on the methods described in Ref.~\cite{EGAM-2018-01}, while corrections for muons follow those in Ref.~\cite{MUON-2018-03}.
Scale factors corresponding to electron reconstruction, muon quality and muon track-to-vertex-matching, derived in 13~\TeV $pp$ collisions, are applied.
Additional correction factors for electron and muon identification, isolation and trigger requirements are estimated by using this \PbPb\ data sample and the tag-and-probe method in $Z \rightarrow \ell\ell$ events~\cite{EGAM-2018-01, MUON-2018-03}.
The detector response corrections are validated using the invariant mass spectrum of $Z\rightarrow \ell\ell$ candidate events in data and simulation.
The events are selected by requiring same-flavor oppositely charged leptons with an invariant mass between 66 and 116~\GeV.

Two signal regions (SR$_{1}$ and SR$_{2}$) are built from events with exactly one muon and one oppositely charged electron. The dilepton invariant mass ($m_{e\mu}$) is required to be $\geq 30$~\GeV to minimize the fake-lepton background contribution. At least two jets with $\pT \geq 35$~\GeV are required. SR$_{1}$ and SR$_{2}$ are defined with a selection on the dilepton transverse momentum $p_\mathrm{T}^{e\mu} > 40$~\GeV and $\leq 40$~\GeV, respectively. Distributions of $m_{e\mu}$ and $p_\mathrm{T}^{e\mu}$ in the $e\mu$ channel can be found in the Supplemental Material~\cite{bib:supp}. Splitting events into two signal regions, where SR$_{1}$ is dominated by signal, improves the measurement sensitivity. The observed number of events in SR$_{1}$ and SR$_{2}$ is 22 and 10, respectively. The main background is the fake-lepton contribution, followed by $Z$+jets with $Z \rightarrow \tau\tau$ and leptonically decaying $\tau$-leptons.

Effects of jet quenching~\cite{Qin:2015srf} are not modeled in MC simulation.
Jet quenching is extensively explored at the LHC~\cite{Cunqueiro:2021wls}, typically reducing the energy scale and broadening the resolution between the final state parton kinematics and the reconstructed jet energy in a HI collision.
Any modifications to the jet spectra may potentially affect the event yields in the signal regions.
Therefore, the detector response and the jet modeling is studied using $Z \rightarrow \ell\ell$ events, with at least one reconstructed jet. Effective correction factors for the jet energy scale are derived separately for central and peripheral collisions to account for the remaining mis-modeling of the detector response, jet-quenching effects, the description of jets from the UE, and random energy fluctuations that are reconstructed as jets.
Moreover, the jet selection in the analysis is optimized such that the impact of the lack of jet-quenching modeling in MC simulation is minimized. Only a requirement on the number of jets is imposed and jet kinematics are not used in the discriminating variable for the cross section measurement.

Fake-lepton backgrounds are from non-prompt leptons, hadrons, and photons that meet the lepton selection criteria.
The main source of fake-lepton background in the dilepton channel is the $W$+jets process.
The normalization and $m_{e\mu}$ distribution shape of the fake-lepton background, which are used in the fitting procedure, are directly estimated from data using the $ABCD$ method. %
Four independent regions are defined: signal region $A$ (corresponding to the combined SR$_{1}$ and SR$_{2}$), and background control region $B$ which is defined using the same selection criteria as in the signal, but requiring two same-sign electrically charged leptons.
Regions $C$ and $D$ are equivalent to the regions $A$ and $B$, respectively, but with inverted lepton isolation requirements.
Regions $B-D$ are therefore enhanced with fake-lepton contributions.
After subtracting potential signal and other background contributions in the regions $B-D$, and assuming no correlations between isolation and the charge requirements, the fake-background contribution in the signal region can be estimated as $N_B N_C/N_D$, where $N_i$ denotes the number of events in the control region $i$.
The shape of the $m_{e\mu}$ distribution of the fake-lepton contribution is estimated by using the region $C$.
Due to the limited number of events in the regions $B-D$, the fake-lepton background is estimated inclusively for electrons and muons. Approximately 36\% and 64\% of the fake-lepton contribution originates from fake electrons and muons, respectively.

There are systematic uncertainties affecting the measurement from the reconstruction of leptons and jets, the fake-lepton background estimate, the signal and background modeling, and the luminosity measurement.
The modeling systematic uncertainties are derived from one-sided variations and symmetrized. Differences between alternative signal MC samples and the nominal one are as much as 12\%.
Uncertainties due to initial-state radiation~(ISR) and final-state radiation~(FSR) are negligible~\cite{TOPQ-2023-32}.
Uncertainties in the electron and muon reconstruction performance are evaluated by varying the corresponding data-to-MC correction factors by their uncertainties.
The jet-related uncertainties are derived from in situ studies of the calorimeter response~\cite{JETM-2018-05}, and from comparisons of the simulated response in samples from different generators.
Uncertainties in the effective jet correction factors are derived from variations with no corrections applied, and symmetrized.

The $VV$ background normalization is allowed to vary by 26\%, which corresponds to the uncertainty in \sherpa\ generator predictions extrapolated to the center-of-mass energy of the \PbPb\ system~\cite{STDM-2012-24, STDM-2016-14}.
The MC prediction of $VV$ production is found to be consistent with the data in a control region, using events that satisfy the signal selection but with zero jets with $\pT>35$~\GeV.
Therefore, no additional shape-related uncertainties are applied to the $VV$ process.
The normalization uncertainty of the $Z\rightarrow \tau\tau$ process is 10\%, reflecting the measurement uncertainty of $Z$ boson production in HI collisions~\cite{HION-2018-10}.
An uncertainty of 9.5\% is applied for the normalization of the $tW$ process~\cite{TOPQ-2018-40}.

The fake-lepton background uncertainties include statistical and systematic variations in both the $m_{e\mu}$ distribution shape and normalization.
Systematic variations in the normalization are evaluated by varying the isolation requirements in regions $C$ and $D$.
The $m_{e\mu}$ distribution shape variation is evaluated using the same-sign electrically charged selection, which corresponds to the combined region $B$ and $D$.
The nominal and the systematically varied fake-lepton $m_{e\mu}$ distribution shapes are smoothed using an exponential function because of the small number of events.

The uncertainty in the integrated luminosity of the combined data sample is 1.5\%. It is derived following a method similar to the one given in Ref.~\cite{DAPR-2021-01}, and using the LUCID-2 detector for the baseline luminosity measurements~\cite{LUCID2}.
\begin{figure}[t!] %
\centering
\subfigure[]{\includegraphics[width=0.35\textwidth]{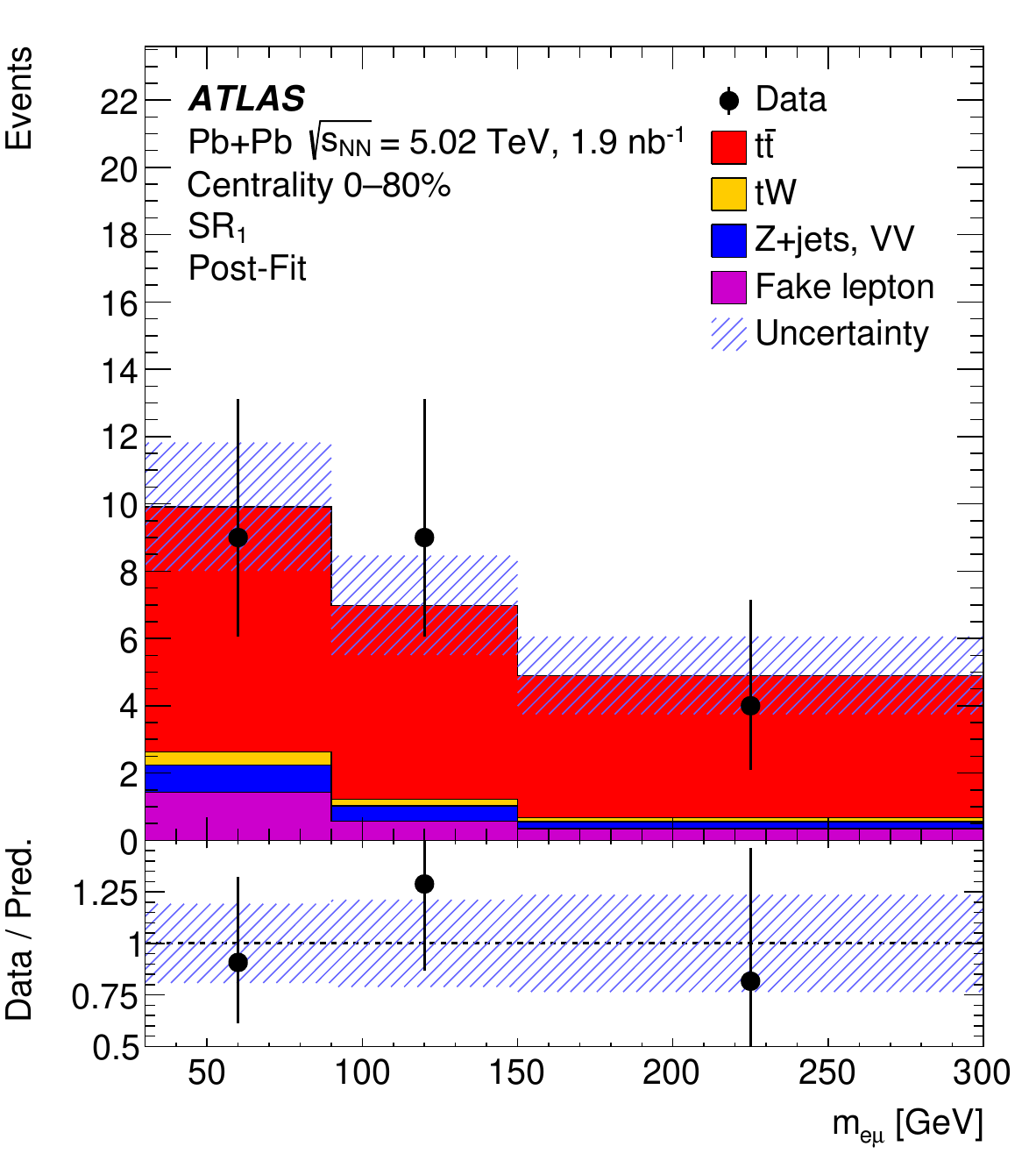}}
\quad
\subfigure[]{\includegraphics[width=0.35\textwidth]{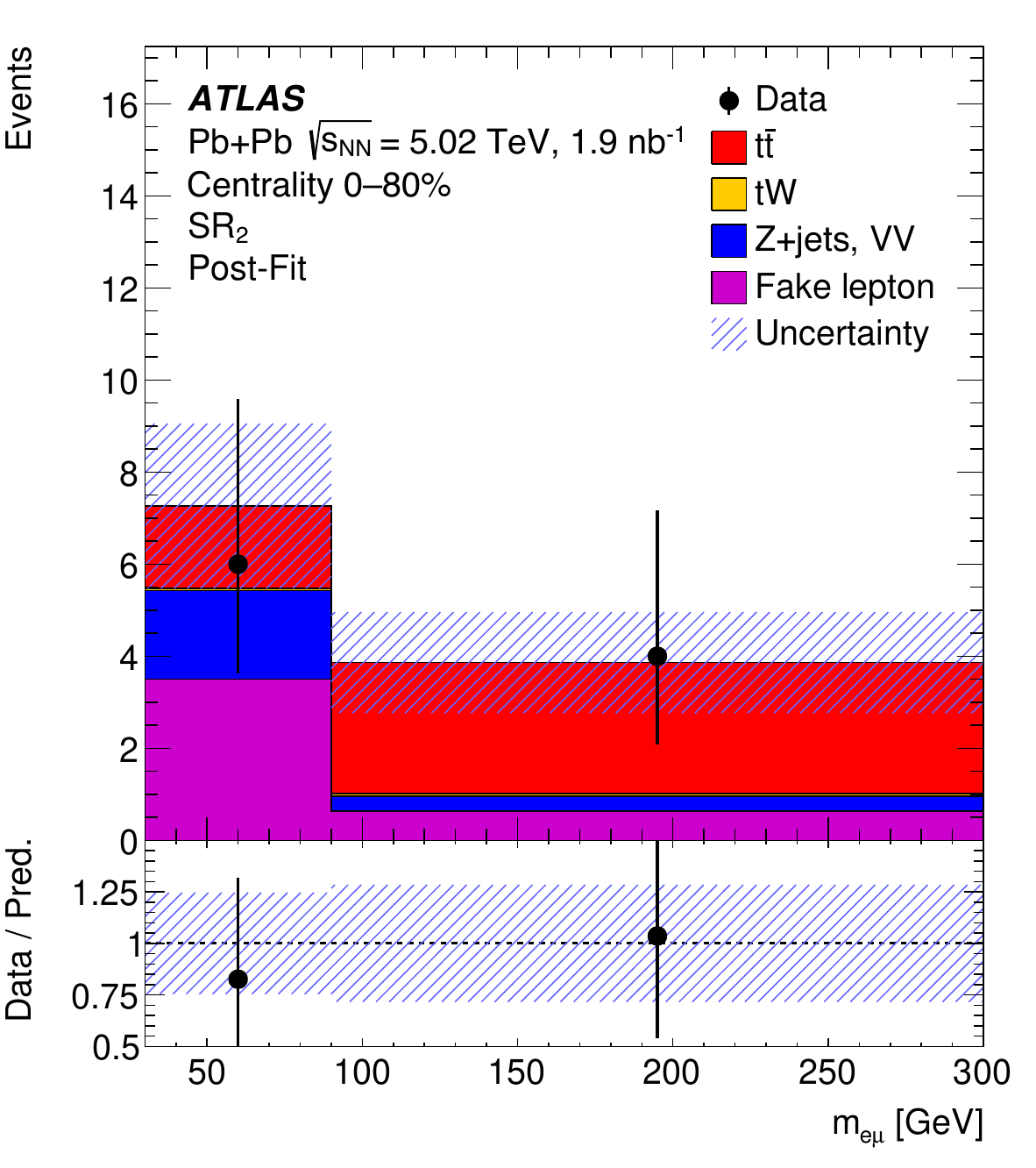}}
\caption{Post-fit distributions of the $m_{e\mu}$ variable in (a)~the SR$_{1}$ and (b)~the SR$_{2}$, with total post-fit uncertainties represented by the hatched area. The filled markers in the
bottom panels show the ratio between data and the sum of predictions.}
\label{fig:data_postfit_full}
\end{figure}

The signal strength $\mu_{\ttbar} \equiv \sigma_{\ttbar}/\sigma^\mathrm{th}_{\ttbar}$ is defined as the ratio of the observed \ttbar\ cross section for the $e\mu$ final state to the SM expectation without any nuclear modifications to the PDFs. The parameter \mutt{} is determined by a combined fit to the $m_{e\mu}$ distribution in both the signal regions. The invariant mass distributions predicted by the fit~(post-fit) in both the signal regions are shown in Figure~\ref{fig:data_postfit_full}. The fit favors smaller background contributions than predicted, leading to a higher signal yield after the fit. The distributions predicted by the fit and the observed distributions are in good agreement.

Systematic uncertainties are included in the fit via nuisance parameters~(NPs), which are additional fit parameters constrained by a Gaussian-distributed probability density function. The breakdown of relative systematic
uncertainties is provided in the Supplemental Material~\cite{bib:supp}.
The leading contributions to the total systematic uncertainty are from the signal modeling, followed by the jet-related uncertainties.
The total relative systematic uncertainty amounts to 18\%, while the statistical uncertainty is 26\%.

The measured \mutt{} value is translated to an inclusive $t\bar{t}$ production cross section using the theoretical prediction for $\sigma^\mathrm{th}_{\ttbar}$ derived in the 0--100\% centrality range at the NNLO+NNLL precision using the \toppp~v2 program. The cross section in the 0--100\% centrality interval is predicted to be 0.19\% higher compared to the 0--80\% range.
The theoretical cross section in $pp$ collisions at $\sqrt{s}=5.02$~\TeV, scaled by the lead mass number~($A_\mathrm{Pb}=208$) squared, is $2.95^{+0.08}_{-0.10}\;\mathrm{(scale)}\;^{+0.10}_{-0.09}\;{(m_t)}\;^{+0.21}_{-0.21}\;\mathrm{(PDF+\alpha_s)}~\mathrm{\mu b}$~\cite{TOPQ-2018-40}.
The measured inclusive $\ttbar{}$ cross section in \PbPb\ collisions is $\sigma_{t\bar{t}} = 3.6\;^{+1.0}_{-0.9}\;\mathrm{(stat.)}\;^{+0.8}_{-0.5}\;\mathrm{(syst.)}~\mathrm{\mu b} = 3.6\;^{+1.2}_{-1.0}\;\mathrm{(tot.)}~\mathrm{\mu b}$.
The total relative uncertainty is 31\%.
The probability for the background-only hypothesis to result in a signal-like excess at least as large as seen in data is derived using the profile-likelihood ratio following the $CL_{S}$ procedure described in Ref.~\cite{Read:2002hq}.
From this, the significance of the observed signal is 5.0 standard deviations, while 4.1 standard deviations are expected.

\begin{figure}[t!]
\centering
\includegraphics[width=0.57\textwidth]{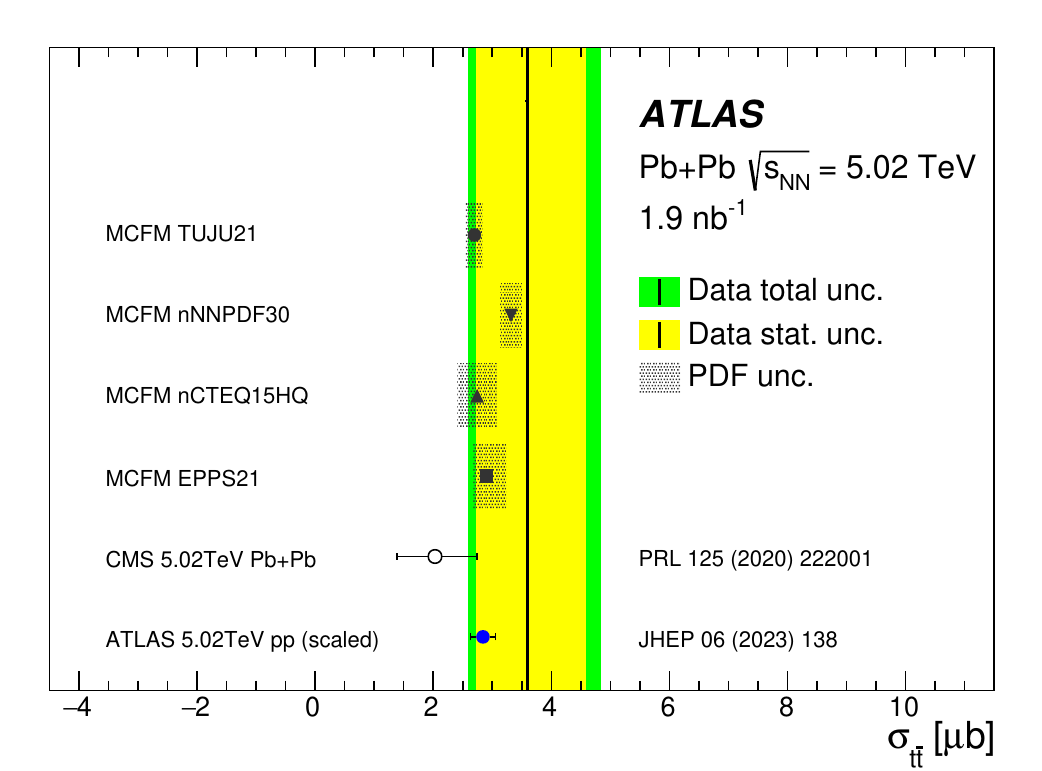}
\caption{Comparison between observed and predicted values of \xsectt{}. This measurement is compared with the CMS Collaboration measurement of \xsectt{} in \PbPb\ collisions at $\sqrt{s_\mathrm{NN}}$=5.02~\TeV~\cite{CMS-HIN-19-001} and the ATLAS measurement in $pp$ collisions at $\sqrt{s}=5.02$~\TeV~\cite{TOPQ-2018-40}. The latter is scaled by $A_\mathrm{Pb}^2$. The theoretical cross section is obtained from NLO calculations using the \mcfm\ generator~\cite{Campbell:2016dks} scaled to the NNLO+NNLL precision by the $K$-factor of $K=1.149$. Predictions are scaled to the \PbPb\ system by $A_\mathrm{Pb}^2$ and given for different nPDF sets. The uncertainty in predictions represents the internal PDF uncertainty.
\label{fig:xsec}}
\end{figure}

Figure~\ref{fig:xsec} shows the measured \xsectt{} compared with the measurement in \PbPb\ collisions at $\sqrt{s_\mathrm{NN}}=5.02$~\TeV by CMS~\cite{CMS-HIN-19-001} and the ATLAS measurement in $pp$ collisions at $\sqrt{s}=5.02$~\TeV~\cite{TOPQ-2018-40}, scaled by $A_\mathrm{Pb}^2$.
Experimental results agree within their uncertainties. The measured cross section is also compared with NLO calculations obtained with the \mcfm\ generator~\cite{Campbell:2016dks} scaled to the NNLO+NNLL precision in QCD using the $K$-factor of $K=1.149$, and scaled to the \PbPb\ system by $A_\mathrm{Pb}^2$.
Four state-of-the-art nPDF sets are used as input~\cite{Klasen:2023uqj} to the \mcfm\ calculations: EPPS21~\cite{Eskola:2021mjl}, nCTEQ15HQ~\cite{Kovarik:2015cma,Duwentaster:2022kpv}, nNNPDF30~\cite{AbdulKhalek:2022fyi} and TUJU21~\cite{Helenius:2022rmf}.
Good agreement is found between the measurement of \xsectt{} and all predictions. Further improvement in the measurement precision is needed to constrain nPDFs.

\FloatBarrier

In conclusion, top-quark pair production in \PbPb\ collisions is observed at the center-of-mass energy of $\sqrt{s_\mathrm{NN}}=5.02$~\TeV per nucleon pair in the ATLAS experiment.
The top-quark pairs are reconstructed using dilepton final states with electrons and muons.
The top-quark pair production cross section is measured to be $\sigma_{t\bar{t}} = 3.6\;^{+1.0}_{-0.9}\;\mathrm{(stat.)}\;^{+0.8}_{-0.5}\;\mathrm{(syst.)} ~\mathrm{\mu b}$
with an observed~(expected) significance of 5.0~(4.1) standard deviations.
The total integrated cross section is measured with a total relative uncertainty of 31\%, dominated by statistical uncertainties in the limited data sample size.
A good agreement is found between the measurement and SM predictions with nuclear effects. The observation of this process consolidates the evidence of the presence of all quark flavors in the pre-equilibrium stage of the QGP, similar to conditions in the early universe~\cite{Heinz:2013th}. This result paves the way for further studies of the QGP and the physics of the early universe with top quarks.

\section*{Acknowledgements}
%

% The next lines are included from the .//acknowledgements/Acknowledgements.tex input file
%
%

%
%

We thank CERN for the very successful operation of the LHC and its injectors, as well as the support staff at
CERN and at our institutions worldwide without whom ATLAS could not be operated efficiently.

The crucial computing support from all WLCG partners is acknowledged gratefully, in particular from CERN, the ATLAS Tier-1 facilities at TRIUMF/SFU (Canada), NDGF (Denmark, Norway, Sweden), CC-IN2P3 (France), KIT/GridKA (Germany), INFN-CNAF (Italy), NL-T1 (Netherlands), PIC (Spain), RAL (UK) and BNL (USA), the Tier-2 facilities worldwide and large non-WLCG resource providers. Major contributors of computing resources are listed in Ref.~\cite{ATL-SOFT-PUB-2025-001}.

We gratefully acknowledge the support of ANPCyT, Argentina; YerPhI, Armenia; ARC, Australia; BMWFW and FWF, Austria; ANAS, Azerbaijan; CNPq and FAPESP, Brazil; NSERC, NRC and CFI, Canada; CERN; ANID, Chile; CAS, MOST and NSFC, China; Minciencias, Colombia; MEYS CR, Czech Republic; DNRF and DNSRC, Denmark; IN2P3-CNRS and CEA-DRF/IRFU, France; SRNSFG, Georgia; BMBF, HGF and MPG, Germany; GSRI, Greece; RGC and Hong Kong SAR, China; ICHEP and Academy of Sciences and Humanities, Israel; INFN, Italy; MEXT and JSPS, Japan; CNRST, Morocco; NWO, Netherlands; RCN, Norway; MNiSW, Poland; FCT, Portugal; MNE/IFA, Romania; MSTDI, Serbia; MSSR, Slovakia; ARIS and MVZI, Slovenia; DSI/NRF, South Africa; MICIU/AEI, Spain; SRC and Wallenberg Foundation, Sweden; SERI, SNSF and Cantons of Bern and Geneva, Switzerland; NSTC, Taipei; TENMAK, T\"urkiye; STFC/UKRI, United Kingdom; DOE and NSF, United States of America.

Individual groups and members have received support from BCKDF, CANARIE, CRC and DRAC, Canada; CERN-CZ, FORTE and PRIMUS, Czech Republic; COST, ERC, ERDF, Horizon 2020, ICSC-NextGenerationEU and Marie Sk{\l}odowska-Curie Actions, European Union; Investissements d'Avenir Labex, Investissements d'Avenir Idex and ANR, France; DFG and AvH Foundation, Germany; Herakleitos, Thales and Aristeia programmes co-financed by EU-ESF and the Greek NSRF, Greece; BSF-NSF and MINERVA, Israel; NCN and NAWA, Poland; La Caixa Banking Foundation, CERCA Programme Generalitat de Catalunya and PROMETEO and GenT Programmes Generalitat Valenciana, Spain; G\"{o}ran Gustafssons Stiftelse, Sweden; The Royal Society and Leverhulme Trust, United Kingdom.

In addition, individual members wish to acknowledge support from Armenia: Yerevan Physics Institute (FAPERJ); CERN: European Organization for Nuclear Research (CERN DOCT); Chile: Agencia Nacional de Investigaci\'on y Desarrollo (FONDECYT 1230812, FONDECYT 1230987, FONDECYT 1240864); China: Chinese Ministry of Science and Technology (MOST-2023YFA1605700, MOST-2023YFA1609300), National Natural Science Foundation of China (NSFC - 12175119, NSFC 12275265, NSFC-12075060); Czech Republic: Czech Science Foundation (GACR - 24-11373S), Ministry of Education Youth and Sports (FORTE CZ.02.01.01/00/22\_008/0004632), PRIMUS Research Programme (PRIMUS/21/SCI/017); EU: H2020 European Research Council (ERC - 101002463); European Union: European Research Council (ERC - 948254, ERC 101089007, ERC, BARD, 101116429), European Union, Future Artificial Intelligence Research (FAIR-NextGenerationEU PE00000013), Italian Center for High Performance Computing, Big Data and Quantum Computing (ICSC, NextGenerationEU); France: Agence Nationale de la Recherche (ANR-20-CE31-0013, ANR-21-CE31-0013, ANR-21-CE31-0022, ANR-22-EDIR-0002); Germany: Baden-Württemberg Stiftung (BW Stiftung-Postdoc Eliteprogramme), Deutsche Forschungsgemeinschaft (DFG - 469666862, DFG - CR 312/5-2); Italy: Istituto Nazionale di Fisica Nucleare (ICSC, NextGenerationEU), Ministero dell'Università e della Ricerca (PRIN - 20223N7F8K - PNRR M4.C2.1.1); Japan: Japan Society for the Promotion of Science (JSPS KAKENHI JP22H01227, JSPS KAKENHI JP22H04944, JSPS KAKENHI JP22KK0227, JSPS KAKENHI JP23KK0245); Netherlands: Netherlands Organisation for Scientific Research (NWO Veni 2020 - VI.Veni.202.179); Norway: Research Council of Norway (RCN-314472); Poland: Ministry of Science and Higher Education (IDUB AGH, POB8, D4 no 9722), Polish National Agency for Academic Exchange (PPN/PPO/2020/1/00002/U/00001), Polish National Science Centre (NCN 2021/42/E/ST2/00350, NCN OPUS 2023/51/B/ST2/02507, NCN OPUS nr 2022/47/B/ST2/03059, NCN UMO-2019/34/E/ST2/00393, NCN \& H2020 MSCA 945339, UMO-2020/37/B/ST2/01043, UMO-2021/40/C/ST2/00187, UMO-2022/47/O/ST2/00148, UMO-2023/49/B/ST2/04085, UMO-2023/51/B/ST2/00920, UMO-2024/53/N/ST2/00869); Slovenia: Slovenian Research Agency (ARIS grant J1-3010); Spain: Generalitat Valenciana (Artemisa, FEDER, IDIFEDER/2018/048), Ministry of Science and Innovation (MCIN \& NextGenEU PCI2022-135018-2, MICIN \& FEDER PID2021-125273NB, RYC2019-028510-I, RYC2020-030254-I, RYC2021-031273-I, RYC2022-038164-I); Sweden: Carl Trygger Foundation (Carl Trygger Foundation CTS 22:2312), Swedish Research Council (Swedish Research Council 2023-04654, VR 2018-00482, VR 2022-03845, VR 2022-04683, VR 2023-03403, VR grant 2021-03651), Knut and Alice Wallenberg Foundation (KAW 2018.0458, KAW 2019.0447, KAW 2022.0358); Switzerland: Swiss National Science Foundation (SNSF - PCEFP2\_194658); United Kingdom: Leverhulme Trust (Leverhulme Trust RPG-2020-004), Royal Society (NIF-R1-231091); United States of America: U.S. Department of Energy (ECA DE-AC02-76SF00515), Neubauer Family Foundation.

%
%

% End of text imported from the .//acknowledgements/Acknowledgements.tex input file

%

%
%
%

%
%
%
%

%
%
%

%
%
%
\clearpage
\printbibliography
\clearpage
% ATLAS Collaboration author list
% Reference date of HION-2022-10 is 2024-09-05
% Author list last updated on date 11-FEB-25
% Data extracted on 11-Feb-2025 for paper reference HION-2022-10
% at 11:47am
 
\begin{flushleft}
\hypersetup{urlcolor=black}
{\Large The ATLAS Collaboration}

\bigskip

\AtlasOrcid[0000-0002-6665-4934]{G.~Aad}$^\textrm{\scriptsize 105}$,
\AtlasOrcid[0000-0001-7616-1554]{E.~Aakvaag}$^\textrm{\scriptsize 17}$,
\AtlasOrcid[0000-0002-5888-2734]{B.~Abbott}$^\textrm{\scriptsize 124}$,
\AtlasOrcid[0000-0002-0287-5869]{S.~Abdelhameed}$^\textrm{\scriptsize 120a}$,
\AtlasOrcid[0000-0002-1002-1652]{K.~Abeling}$^\textrm{\scriptsize 57}$,
\AtlasOrcid[0000-0001-5763-2760]{N.J.~Abicht}$^\textrm{\scriptsize 51}$,
\AtlasOrcid[0000-0002-8496-9294]{S.H.~Abidi}$^\textrm{\scriptsize 30}$,
\AtlasOrcid[0009-0003-6578-220X]{M.~Aboelela}$^\textrm{\scriptsize 46}$,
\AtlasOrcid[0000-0002-9987-2292]{A.~Aboulhorma}$^\textrm{\scriptsize 36e}$,
\AtlasOrcid[0000-0001-5329-6640]{H.~Abramowicz}$^\textrm{\scriptsize 156}$
% ...
\bigskip
\\

% ...
$^{17}$Department for Physics and Technology, University of Bergen, Bergen; Norway.\\
% ...
$^{30}$Physics Department, Brookhaven National Laboratory, Upton NY; United States of America.\\
% ...
$^{36}$$^{(a)}$Facult\'e des Sciences Ain Chock, Universit\'e Hassan II de Casablanca;$^{(b)}$Facult\'{e} des Sciences, Universit\'{e} Ibn-Tofail, K\'{e}nitra;$^{(c)}$Facult\'e des Sciences Semlalia, Universit\'e Cadi Ayyad, LPHEA-Marrakech;$^{(d)}$LPMR, Facult\'e des Sciences, Universit\'e Mohamed Premier, Oujda;$^{(e)}$Facult\'e des sciences, Universit\'e Mohammed V, Rabat;$^{(f)}$Institute of Applied Physics, Mohammed VI Polytechnic University, Ben Guerir; Morocco.\\
% ...
$^{46}$Physics Department, Southern Methodist University, Dallas TX; United States of America.\\
% ...
$^{51}$Fakult\"{a}t Physik , Technische Universit{\"a}t Dortmund, Dortmund; Germany.\\
% ...
$^{57}$II. Physikalisches Institut, Georg-August-Universit\"{a}t G\"ottingen, G\"ottingen; Germany.\\
% ...
$^{105}$CPPM, Aix-Marseille Universit\'e, CNRS/IN2P3, Marseille; France.\\
% ...
$^{120}$$^{(a)}$New York University Abu Dhabi, Abu Dhabi;$^{(b)}$United Arab Emirates University, Al Ain; United Arab Emirates.\\
% ...
$^{124}$Homer L. Dodge Department of Physics and Astronomy, University of Oklahoma, Norman OK; United States of America.\\
% ...
$^{156}$Raymond and Beverly Sackler School of Physics and Astronomy, Tel Aviv University, Tel Aviv; Israel.\\
% ...
$^{a}$ Also Affiliated with an institute covered by a cooperation agreement with CERN.\\
$^{b}$ Also at An-Najah National University, Nablus; Palestine.\\
$^{c}$ Also at Borough of Manhattan Community College, City University of New York, New York NY; United States of America.\\
$^{d}$ Also at Center for High Energy Physics, Peking University; China.\\
$^{e}$ Also at Center for Interdisciplinary Research and Innovation (CIRI-AUTH), Thessaloniki; Greece.\\
$^{f}$ Also at CERN, Geneva; Switzerland.\\
% ...

\end{flushleft}

% Created with Glance <Atlas.Glance@cern.ch>

%

%
%
%
%

%
%

%

%
%
%
%

\end{document}